\documentclass[manuscript, nonacm]{acmart}

\usepackage{xcolor}
\usepackage[utf8]{inputenc} 
\usepackage[T1]{fontenc}   
\usepackage{hyperref}      
\usepackage{url}           
\usepackage{booktabs}       
\usepackage{amsfonts}       
\usepackage{nicefrac}      
\usepackage{microtype}    
\usepackage{bm}
\usepackage{courier}
\usepackage{soul}
\usepackage{graphicx}
\usepackage{xspace}
\usepackage{subcaption}
\usepackage{colortbl}
\usepackage{flafter}
\usepackage[utf8]{inputenc}
\usepackage[most]{tcolorbox}
\usepackage{makecell}

\usepackage{blindtext}
\usepackage{etoolbox}

\providetoggle{showcomments}
\settoggle{showcomments}{true}
\iftoggle{showcomments}{% Show with comments and old text
    \newcommand{\resolved}[3][]{\ifstrequal{#1}{resolved}{\textcolor{blue}{RESOLVED:}~\textbf{{\MakeUppercase #2:}}~{#3}}{\textbf{\MakeUppercase #2:}~#3}}
    \newcommand{\lisa}[2][]{\textcolor{olive}{\resolved[#1]{lisa}{#2}}}
    \newcommand{\roy}[2][]{\textcolor{red}{\resolved[#1]{roy}{#2}}}
    \newcommand{\abbas}[2][]{\textcolor{violet}{\resolved[#1]{abbas}{#2}}}
    \newcommand{\sina}[2][]{\textcolor{magenta}{\resolved[#1]{sina}{#2}}}
    \newcommand{\sara}[2][]{\textcolor{orange}{\resolved[#1]{sara}{#2}}}
    \newcommand{\shayan}[2][]{\textcolor{blue}{\resolved[#1]{shayan}{#2}}}
    
    \newcommand{\todo}[1]{\textcolor{blue}{\textbf{TODO:} #1}}
}{% Show final version
    
    \newcommand{\todo}[1]{}
    \newcommand{\lisa}[2][]{}
    \newcommand{\roy}[2][]{}
    \newcommand{\abbas}[2][]{}
    \newcommand{\sina}[2][]{}
    \newcommand{\sara}[2][]{}
    \newcommand{\shayan}[2][]{}
    
}

\makeatletter
\let\@authorsaddresses\@empty
\makeatother

\AtBeginDocument{%
  \providecommand\BibTeX{{%
    \normalfont B\kern-0.5em{\scshape i\kern-0.25em b}\kern-0.8em\TeX}}}

\makeatletter
\patchcmd{\maketitle}{\@copyrightspace}{}{}{}
\makeatother

\newcommand\blfootnote[1]{
    \begingroup
    \renewcommand\thefootnote{}\footnote{#1}
    \addtocounter{footnote}{-1}
    \endgroup
}

\begin{document}

%%%%%%%%%%%%%%%%%%%%%%%%%   Title   %%%%%%%%%%%%%%%%%%%%%%%%%%%
\title[From Words and Exercises to Wellness: Farsi Chatbot for Self-Attachment Technique]{From Words and Exercises to Wellness: Farsi Chatbot for Self-Attachment Technique}

%%%%%%%%%%%%%%%%%%%%%%%%%   Authors   %%%%%%%%%%%%%%%%%%%%%%%%%%%
\author{Sina Elahimanesh}
\email{sina.elahimanesh@sharif.edu}
\orcid{0000-0001-7251-6661}
\affiliation{%
  \institution{Sharif University of Technology}
  \city{Tehran}
  \country{Iran}
}

\author{Shayan Salehi}
\email{shayan.salehi81@sharif.edu}
\affiliation{%
  \institution{Sharif University of Technology}
  \city{Tehran}
  \country{Iran}
}

\author{Sara Zahedi Movahed}
\email{sara.zahedi@sharif.edu}
\affiliation{%
  \institution{Sharif University of Technology}
  \city{Tehran}
  \country{Iran}
}

\author{Lisa Alazraki}
\affiliation{%
  \institution{Imperial College London}
  \city{London}
  \country{United Kingdom}
}

\author{Ruoyu Hu}
\affiliation{%
  \institution{Imperial College London}
  \city{London}
  \country{United Kingdom}
}

\author{Abbas Edalat}
\email{a.edalat@imperial.ac.uk}
\affiliation{%
  \institution{Imperial College London}
  \city{London}
  \country{United Kingdom}
}

\renewcommand{\shortauthors}{Elahimanesh, et al.}

\blfootnote{Corresponding authors: Sina Elahimanesh, sina.elahimanesh@sharif.edu, Sharif University of Technology, Tehran, Iran. Shayan Salehi, shayan.salehi81@sharif.edu, Sharif University of Technology, Tehran, Iran. Sara Zahedi Movahed, sara.zahedi@sharif.edu, Sharif University of Technology, Tehran, Iran. Abbas Edalat, a.edalat@imperial.ac.uk, Imperial College London, London, United Kingdom.}

%%%%%%%%%%%%%%%%%%%%%%%%%   Abstract   %%%%%%%%%%%%%%%%%%%%%%%%%%%
\begin{abstract}

In this work, we develop a cross-platform, voice-capable chatbot in Farsi to guide users through Self-Attachment (SAT), a novel, self-administered, holistic psychological technique based on attachment theory. Our chatbot uses rule-based and classification-based modules for user comprehension and navigates a dialogue flowchart accordingly, recommending appropriate SAT exercises. For this purpose, we develop a 12-class emotion recognition module with 
accuracy above 92\%. 
The chatbot keeps the conversation engaging by retrieving responses from a large dataset created with the aid of ParsGPT-2 and a reinforcement learning approach. Our chatbot also offers a question-answering module suitable for first-time users called SAT Teacher. We evaluate our platform in a non-clinical study with N=51 volunteers and over 2,000 dialogues with the chatbot. The results indicate that the platform was engaging to most users (76\%), 73\% felt better after the interactions, and 74\% were satisfied with the SAT Teacher's performance. 

\end{abstract}

%%%%%%%%%%%%%%%%%%%%%%%%%   CCS Concepts   %%%%%%%%%%%%%%%%%%%%%%%%%%%


%%%%%%%%%%%%%%%%%%%%%%%%%   Keywords   %%%%%%%%%%%%%%%%%%%%%%%%%%%
\maketitle

%%%%%%%%%%%%%%%%%%%%%%%%%   Introduction   %%%%%%%%%%%%%%%%%%%%%%%%%%%
\section{Introduction}
According to the World Health Organization, the prevalence of anxiety and depression increased by 25\% globally after the COVID-19 pandemic \cite{who2022}. Iran was among the countries most affected by COVID-19 within the Middle Eastern region \cite{lancet2022, BALOCH20201247}, suffering high infection and mortality rates \cite{ghafari2021, ebrahimoghli2023}. Previous research has recorded increased levels of anxiety and depression across different age and socio-economic groups within Iran, attributable to the trauma of the pandemic and the social isolation resulting from prolonged quarantine periods \cite{maroufizadeh2022, MOGHANIBASHIMANSOURIEH2020102076, sabouhi2022, shahriarirad2021}.

In this paper, we develop a Farsi chatbot to enhance mental health. Digital technologies have been widely investigated as a potential tool for supplementing traditional therapy sessions~\cite{weightman2020}, notably in middle-income countries where access to a mental health professional may not always be possible~\cite{Fu2020}. Chatbots, in particular, have been proven effective at easing mental disorders ~\cite{islam2021mobile, He2022} and have been shown to achieve high levels of user satisfaction~\cite{cameron2019assessing, 10.1145/3485874, shah2022}. As a leading example, the Woebot platform, which delivers Cognitive Behavioural Therapy (CBT), has been available and evaluated since 2017 for English-speaking populations around the world~\cite{fitzpatrick2017delivering}. To our knowledge, this is the first chatbot for mental health in Persian, also called Farsi, the official language in Iran and spoken by tens of millions of other people in the Middle East~\cite{windfuhr2009iranian}. 
A Farsi-speaking chatbot could also aid mental health interventions in Afghanistan, where Dari, sometimes called Farsi, is the most widely spoken language. Farsi and Dari are dialects of each other and are mutually intelligible languages in written form. The incidence of anxiety and depression in Afghanistan has risen during the pandemic 
\cite{rasib2021, niazi2022}, and the country lacks 
appropriate therapeutic infrastructure to address the issue \cite{trani2016}. Farsi is also spoken in Tajikistan, some of the northern and western parts of Pakistan, and the southern parts of the Persian Gulf~\cite{spooner20124}.

Our dialogue agent is grounded in Self-Attachment Technique (SAT)~\cite{edalat2015introduction, Edalat2017}, a recently developed psychotherapy framework based on attachment theory, a main paradigm in developmental psychology, and several interdisciplinary concepts such as neuroscience, neuroplasticity, mathematical models, and artificial neural networks \cite{Edalat2017}. This technique helps the user create first a compassionate connection and then an internal affectional bond between their ‘adult self' (the logical and reasoning self) and their ‘childhood self' (the emotional self), as represented by the user's favorite childhood photo or the avatar created from this photo. This scheme aims to optimize emotion regulation circuits. By positively interacting with the childhood self, the technique moderates arousal levels, minimizing negative emotions like anxiety and depression and maximizing positive emotions. It is possible to capture the impact of Self-Attachment protocols with a mathematical model based on the Hopfield network, one of the first artificial models of associative memory~\cite{edalat2015introduction}. There are also computational models of the brain that support the SAT protocol~\cite{cittern2017neural,cittern2018intrinsic}.

SAT has proven effective in both clinical and non-clinical populations \cite{ijerph19116376, polydorou2021}, and it has been shown to have a considerable effect size in a three-month pilot study in a population of chronically depressed and anxious women in Iran~\cite{edalat2022pilot}. 
% SAT consists of 26 self-administered exercises for the user to enhance their emotional self-regulation and well-being capacity. 
SAT consists of 26 self-administered exercises for users to enhance their emotional self-regulation and well-being.
Due to its self-administered nature, SAT lends itself well to being delivered digitally~\cite{ghaznavi2019usability, polydorou2021interactive, edalat2023VRSAT}. Alazraki et al. present a new dataset and a computational strategy for a digital coach to guide experimenters in practicing SAT. They identify underlying emotions and craft human-like characters to achieve a high level of engagement in the virtual therapy sessions \cite{9750315}.
In addition, a VR platform for SAT is developed and evaluated by asking participants to rate the platform's level of immersion, learnability, and overall usability, with promising results~\cite{ghaznavi2019usability}. A high-end implementation of this VR platform, using Oculus Quest, provides suggestions of appropriate SAT exercises according to the user's sentiment~\cite{polydorou2021interactive}. A low-end platform implementation using Google Cardboard has shown large effect sizes for enhancing compassion and well-being in an 8-week pilot study in the non-clinical population~\cite{edalat2023VRSAT}.

Our work extends an existing text-only SAT chatbot in English \cite{9750315} by creating a large pool of Farsi utterances with a reinforcement learning approach using ParsGPT-2 \cite{ParsGPT2}. This approach is similar to the one explored in \cite{10063528} for translating English utterances into Standard Chinese. At each stage of the conversation, we then select the most coherent utterance from the dataset using ParsBERT \cite{Farahani_2021} embeddings. In addition, our chatbot relies on a mixture of model-based and rule-based modules to comprehend user sentiment and intent. Our emotion recognition module, trained on a dataset of over 6,000 examples that we collected, identifies 12 fine-grained emotions with 92.13\% accuracy.

In order to maximize usability, our chatbot supports both text-based and voice-based modalities and is cross-platform, and we include the option for users to change the tone of the conversation from formal to informal. Finally, we augment the platform with a QA system called SAT Teacher to answer user questions about SAT. 
% The chatbot is evaluated via a human study with N=51 volunteers from the non-clinical population. 
We evaluated the chatbot via a human study with N=51 volunteers from the non-clinical population.
We find that the majority of study participants rate the chatbot highly on all metrics (with an overall average score of 7.25 on a 1-10 scale). The performance of the SAT teacher also scores positively (average score 7.42).

% Our main contributions are the following: (1) we developed a Farsi chatbot for enhancing mental health. To the best of our knowledge, ours is the first mental health chatbot in Farsi; (2) we collected monolingual Farsi datasets and made them publicly available on
Our main contributions are: (1) We developed a Farsi chatbot to enhance mental health. To the best of our knowledge, ours is the first mental health chatbot in Farsi; (2) we collected monolingual Farsi datasets and made them publicly available on 
GitHub\footnote{\href{https://github.com/SATProject/Datasets}{https://github.com/SATProject/Datasets}}. Good-quality monolingual datasets are fundamental for moving Farsi from a low to a high-resource language~\cite{shamsfard2019challenges}; (3) we examined the effectiveness of the chatbot in a non-clinical population with N=51 participants, paving the way for future clinical studies applying similar technologies.

%%%%%%%%%%%%%%%%%%%%%%%%%   Related Work   %%%%%%%%%%%%%%%%%%%%%%%%%%%
\section{Related Work}

\subsection{Digital Psychotherapy}
Digital technologies have been touted as a viable alternative to enhance, if not replace altogether, traditional therapy sessions wherever barriers exist to access in-person mental health services, e.g., in rural areas \cite{doi:10.1177/0300060520928686}. Previous work has shown that digital psychotherapy can be as effective at improving mental health as face-to-face therapeutic sessions~\cite{weightman2020digital}, and its effectiveness is consistent across the population independently of gender, self-reported financial status and physical health~\cite{marcelle2019effectiveness}. Moreover, both medical professionals and patients have shown a positive attitude toward digital applications for mental health \cite{mayer2019}, particularly toward chatbots \cite{Abd-Alrazaq2021}.

\subsection{Chatbots in Mental Health Support}
The use of chatbots to facilitate mental healthcare has been widely investigated in the literature, with the technology achieving high levels of user satisfaction~\cite{cameron2019assessing, 10.1145/3485874, shah2022}. Previous work has also explored the effectiveness of chatbots for easing mental disorders during the COVID-19 pandemic, with positive results~\cite{islam2021mobile, He2022}. 

Some scholars have highlighted a preference among youth and young adults toward dialogue agents for mental health support, as opposed to traditional psychotherapy sessions~\cite{info:doi/10.2196/43102, doi:10.1111/ap.12341}. Dialogue agents have proven successful at enhancing mental wellness in the 15-17 age group~\cite{grove2021co}, the 16-21 age group \cite{brandtzaeg2021social}, and a population of university students aged 19-28 \cite{LIU2022100495}. While several human studies with existing chatbots solely focus on younger subsets of the population, it has been shown that this technology is also effective at improving mental health among older age groups \cite{singh2023, hassan2021development, chou2023}. For our study, we hence recruited participants aged 19-65.

\subsection{Farsi Chatbot Technologies}
It should be noted that data scarcity is a major concern when developing dialogue agents in languages other than English~\cite{gerhard-young-etal-2022-low, AHMED2022100057}. Farsi is identified as a low-resource language within the Low Resource Languages for Emergent Incidents (LORELEI) program \cite{strassel-tracey-2016-lorelei}, and scholars have highlighted the difficulty of finding good-quality monolingual datasets in Farsi~\cite{shamsfard2019challenges, arshia2022peqa}. In an effort to move Farsi from a low to a high-resource language, recent years have seen the release of a number of medium-to-large-scale Farsi datasets, particularly focused on conversation and question-answering~\cite{arshia2022peqa, mobasher2023, Suri2021MeDiaQAAQ, 9729745}. Additionally, significant work has been done toward developing task-oriented dialogue systems in Farsi~\cite{borhanifard2020persian, shahedi2023chatparse}. We aim to reinforce this positive trend with the release of our chatbot, as well as the associated public dataset in Farsi, which focuses on fine-grained emotion classification in dialogue.

\begin{figure}[t]
  \centering
  \includegraphics[width=0.85\textwidth]{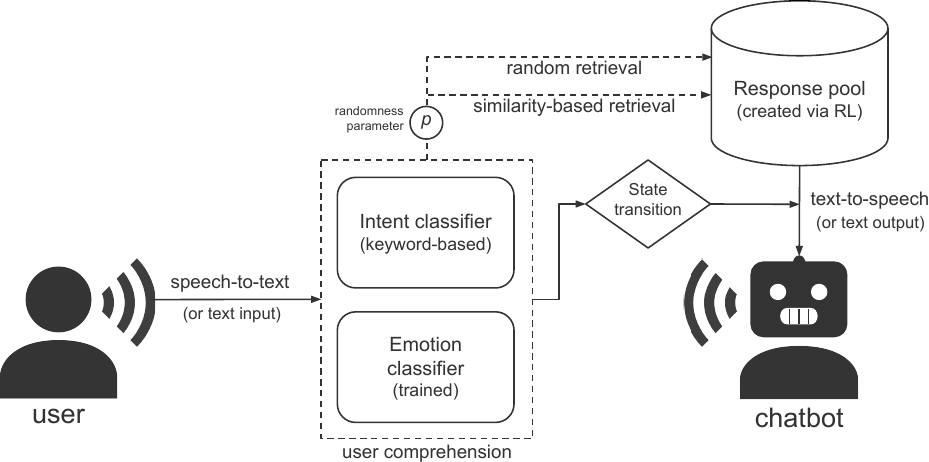}
  \caption{The pipeline of the SAT chatbot.}
  \label{fig:pipeline}
\end{figure}

%%%%%%%%%%%%%%%%%%%%%%%%%   Proposed Method   %%%%%%%%%%%%%%%%%%%%%%%%%%%

\section{Proposed Method}
\begin{figure}[hbt!]
\centering
  \includegraphics[width=0.746\textwidth]{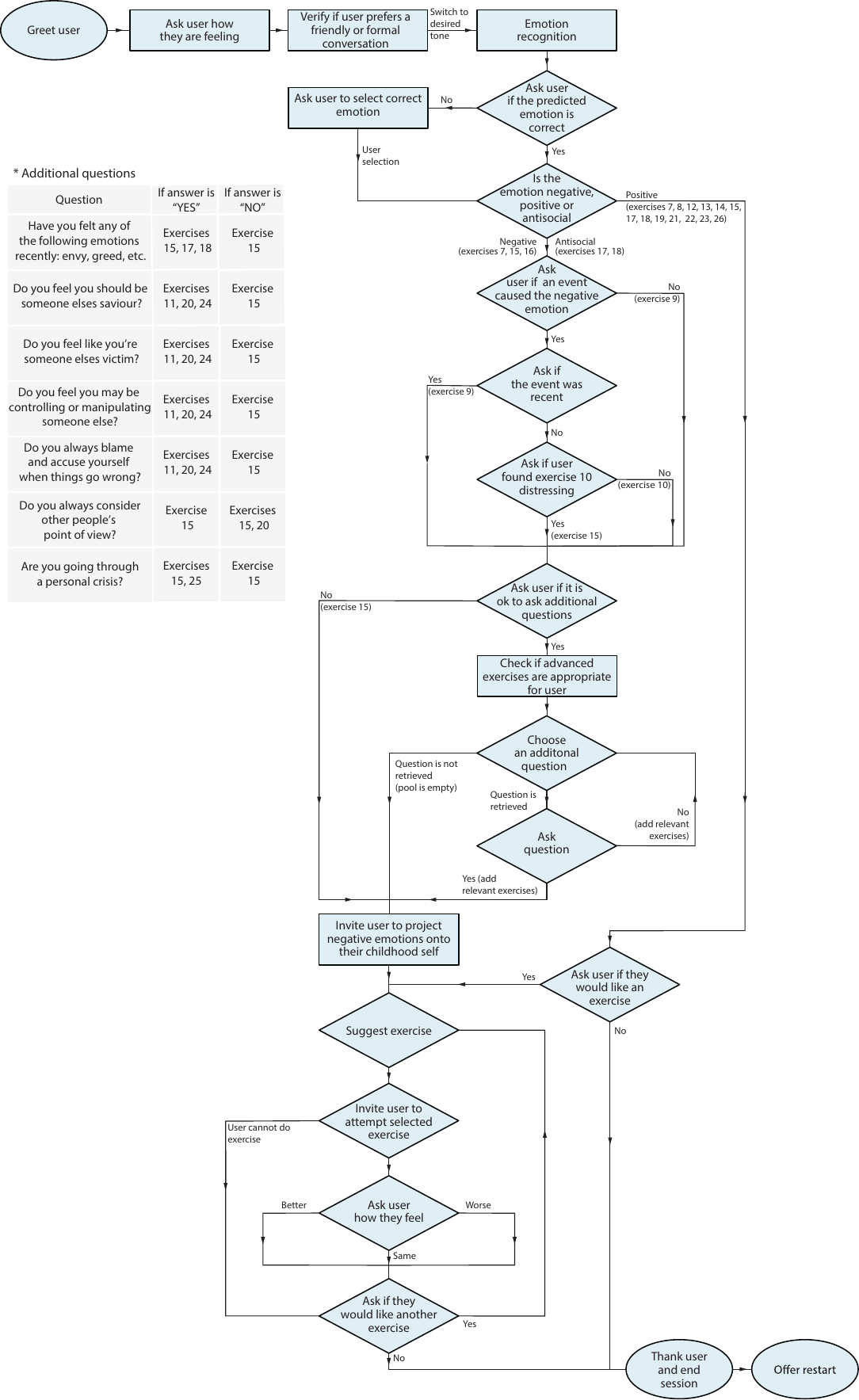}
  \caption{The conversation flowchart of the chatbot, adapted from \cite{9750315}.}
  \label{fig:flowchart}
\end{figure}

Our chatbot follows the same rule-based flowchart as in \cite{9750315}. For each step in the conversation, we implemented user comprehension modules based on detecting keyword combinations (e.g., to determine negative/positive intent) and classification to identify the user's current emotion. A detailed illustration of our flowchart is given in Figure~\ref{fig:flowchart}. Following input comprehension, our framework selects a response from a pool of available utterances based on embedding similarity. The chatbot's conversation pipeline is shown in Figure \ref{fig:pipeline}.

\subsection{Emotion Classification Module}\label{sentiment_analysis}

Fine-grained emotion recognition is a necessary feature for a chatbot offering mental health support. Hence, we implemented an emotion classification module able to identify 12 emotions: \textit{Happy, Angry, Anxious, Ashamed, Disappointed, Disgusted, Envious, Guilty, Insecure, Loving, Sad}, and \textit{Jealous}. Due to the scarcity of Farsi datasets labeled for this purpose, we created a new corpus by prompting GPT-3.5 \cite{openai2022chatgpt} in a multi-step fashion to produce diverse Farsi utterances of varying length for each of the 12 emotions. In-context exemplars were deemed unnecessary for this task and thus were not provided to the model. Since Farsi has specific, common expressions to indicate the speaker's emotion, which were not included in the utterances generated by GPT-3.5, we manually created additional samples, obtaining a final corpus of 6,168 labeled utterances. In Table \ref{tab:emo_data_stats}, we show the distribution of the classes in our dataset for emotion classification in Farsi. As evidenced in the table, the classes are approximately balanced. This dataset was used to finetune a ParsBERT model \cite{Farahani_2021} for emotion classification, with results shown in Table \ref{tab:sentiment_analysis_results}. Our model's performance (92.13\% accuracy, 91.86\% F1-score) surpasses previous similar models \cite{Ali2021emotion, Amir22emotion} while classifying a wider range of emotions, demonstrating significant advancement in the Farsi linguistic domain.

\begin{table}[!ht]
    %\centering
  \caption{Emotion dataset statistics.}
   \setlength{\tabcolsep}{32pt}
   \renewcommand{\arraystretch}{1}
    \begin{tabular}{lccc}
        \toprule
        Emotion       & \# of utterances & \% of corpus \\
        \midrule
        Happy         & 518 & 8.4\% 
        \\
        Angry         & 550 & 8.9\% 
        \\
        Anxious       & 508 & 8.2\%
        \\
        Ashamed       & 487 & 7.9\%  
        \\
        Disappointed  & 520 & 8.4\%
        \\
        Disgusted     & 477 & 7.7\% 
        \\
        Envious       & 498 & 8.1\% 
        \\
        Guilty        & 567 & 9.2\% 
        \\
        Insecure      & 504 & 8.2\%   
        \\
        Loving        & 500 & 8.1\%
        \\
        Sad           & 524 & 8.5\% 
        \\
        Jealous       & 515 & 8.3\% 
        \\
        \cmidrule{1-2}
        Total         & 6,168 & 
        \\
        \bottomrule
    \end{tabular}
    \label{tab:emo_data_stats}
\end{table}

\begin{table}[!ht]
    %\centering
  \caption{Classifier results for each emotion class.}
   \setlength{\tabcolsep}{32pt}
   \renewcommand{\arraystretch}{1.2}
    \begin{tabular}{lccc}
        \toprule
        Emotion       & Precision & Recall & F1-Score \\
        \midrule
        Happy         & 90.21\% & 98.36\% & 94.11\%  \\
        Angry         & 89.38\% & 77.98\% & 82.29\%  \\
        Anxious       & 71.78\% & 86.34\% & 78.39\%  \\
        Ashamed       & 98.22\% & 99.05\% & 98.63\%  \\
        Disappointed  & 94.41\% & 97.08\% & 95.73\%  \\
        Disgusted     & 99.01\% & 97.96\% & 98.48\%  \\
        Envious       & 98.09\% & 99.12\% & 98.60\%  \\
        Guilty        & 97.89\% & 99.16\% & 98.52\%  \\
        Insecure      & 91.45\% & 74.87\% & 82.33\%  \\
        Loving        & 98.21\% & 96.29\% & 97.24\%  \\
        Sad           & 79.68\% & 81.44\% & 80.55\%  \\
        Jealous       & 98.00\% & 99.09\% & 98.54\%  \\
        \midrule
        Accuracy          &         &         & 92.13\%   \\
        Macro average     & 92.19\% & 92.23\% & 92.00\%   \\
        Weighted average  & 92.13\% & 92.13\% & 91.86\%   \\
        \bottomrule
    \end{tabular}
    \label{tab:sentiment_analysis_results}
\end{table}

\subsection{Reinforcement Learning Approach to Building Conversation Pools}

To ensure that our chatbot's responses are retrieved from a pool of candidates that are appropriate for a mental health context, we finetuned a ParsGPT-2 model via proximal policy optimization (PPO) to rewrite utterances in an empathetic manner. This is similar to the approach used in \cite{10063528}, with the addition of a similarity reward during training to ensure that the generated paraphrases remain similar to the input utterance. To create the training data for this model, we first collected a corpus of 5,598 utterances by translating the English-only dataset EmpatheticPersonas \cite{9750315} (hereafter referred to as EP) into Farsi. An initial translation was procured using the publicly accessible Google Translate tool\footnote{https://translate.google.com}, with subsequent automated and manual refinements to enhance fidelity. To increase the diversity of our data, we prompted GPT-3.5 to produce further Farsi utterances for each step of the conversation, using both a formal and an informal tone. We provided the model with 20 translated utterances as in-context exemplars each time. After manual filtering, we obtained a corpus of utterances that was approximately 1.5 times the size of the initial one.

We used this dataset to train ParsGPT-2 to rewrite utterances empathetically while maintaining fluency and semantic similarity to the input. Similarly to previous work \cite{ppo2017, Ziegler19finetune}, we employed two versions of the ParsGPT-2 model: an Active model fine-tuned via PPO and a Reference model whose purpose is to deter abrupt policy alterations by computing during training a KullbackLeibler (KL) penalty $D_{KL}(P \lVert Q)$, where $P$ and $Q$ represent the policy distributions of the Active and Reference models, respectively. Both models were fed the same prompts, ensuring consistency in terms of data and evaluation. Our training reward consisted of the weighted sum of the individual empathy ($r_e$), fluency ($r_f$) and semantic ($r_s$) rewards, analogous to those described in \cite{10063528}, with the addition of a similarity reward ($r_{sim}$) given by the cosine similarity between the ParsBERT embeddings of the original input utterances and those of the rewritings. The latter's inclusion was motivated by an increased tendency of the Farsi model to significantly deviate from the input compared to \cite{10063528}, rendering the semantic reward alone insufficient to ground the generated rewriting to the given context. To compute the empathy and semantic rewards, we used the same classification methods as in \cite{10063528} and fine-tuned an XLM-R model \cite{xlmmodel} on the EP data previously translated into Farsi, achieving strong performance in both cases (92.05\% accuracy and 92.12\% F1-score for the empathy task, 96.89\% accuracy and 96.24\% F1-score for the semantic task). We computed the perplexity score using ParsGPT-2 to calculate the fluency reward. For all three rewards, we refer to \cite{10063528} for detailed descriptions and formulas. The individual rewards were normalized via min-max scaling to take values in the [-1, 1] interval, and the final training reward was given by

\vspace{-5pt}

\begin{equation}
    r = w_f r_f + w_s r_s + w_e r_e + w_{sim} r_{sim}
\end{equation}

\vspace{5pt}

where the weights $w_f$, $w_s$, $w_e$ and $w_{sim}$ were tuned manually. 
Finally, we used the finetuned ParsGPT-2 model to produce empathetic rewritings of each base utterance in the input dataset, obtaining a pool of 152,000 utterances.

\subsection{Response Selection}
At each step in the conversation flow, appropriate responses are selected from the relevant subsection of the utterance pool by computing the cosine similarity between their ParsBERT embeddings and the averaged embeddings of the previous conversation context. While retrieving the utterance most similar to the context guarantees coherence, we found that allowing the chatbot to select a random utterance occasionally has the potential to increase novelty in the conversation. Therefore, we sample random utterances with probability $p$. We set $p=0.5$ during our evaluation study.

\begin{figure}[!b]
  \centering
  \includegraphics[width=1\textwidth]{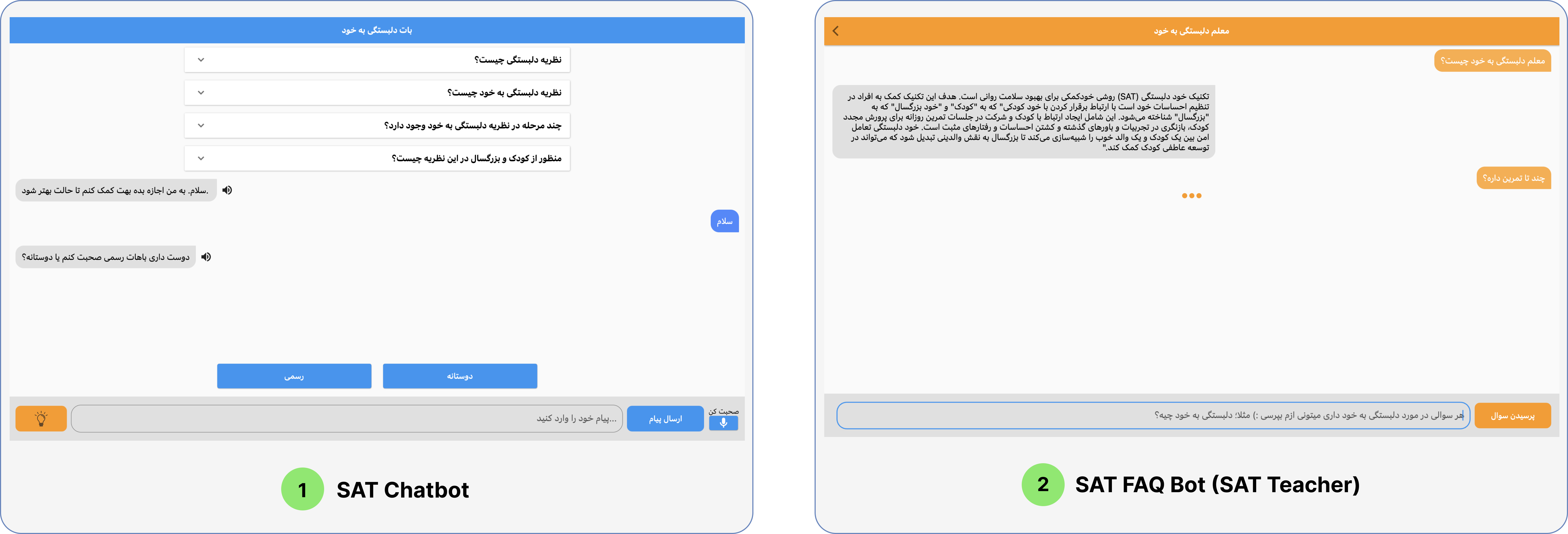}
  \caption{User interface of our platform in Chatbot and SAT Teacher modality.}
  \label{fig:website}
\end{figure}

\subsection{SAT Teacher Module}\label{sat-teacher-sec}
In order to help users familiarize themselves with SAT, we implemented a QA module called SAT Teacher within our chatbot. This module is based on a corpus of 140 QA pairs manually crafted from the scientific literature on SAT. Each pair contains two analogous, paraphrased questions to increase the probability of matching a user's query to the correct answer. In our framework, user queries are matched to the most similar question in the corpus, found by computing the cosine similarity over ParsBERT embeddings, and the corresponding answer is then retrieved. To validate our method, we split a portion of the corpus and crafted new queries from the existing questions, using synonym replacement, paraphrasing, and manually inserting introductory and concluding phrases that are characteristic of the Farsi language. We found that the SAT Teacher was able to return the correct answer in 80.12\% of cases, having observed the new query.

%%%%%%%%%%%%%%%%%%%%%%%%%   Experimental Evaluation and Results   %%%%%%%%%%%%%%%%%%%%%%%%%%%
\section{Experimental Evaluation and Results}

We conducted a user study in a non-clinical population of Farsi speakers to evaluate the feasibility of employing the chatbot in human trials. In this experiment, we primarily followed the guidelines of the UK government for feasibility studies~\cite{noauthor_feasibility_nodate-1} to determine the users' degree of engagement with and acceptability of the chatbot. Additionally, we aimed at evaluating the potential efficacy of the chatbot at interacting empathetically with users---a key factor in any successful therapy~\cite{elliott2018therapist}---as well as correctly detecting the user's emotional state and helping to reduce negative emotions and increase positive ones. The ethical approval for this study was obtained from Imperial College London's Science, Engineering, and Technology Research Ethics Committee.

\subsection{Preliminary Evaluation}
As a first step, we ran a preliminary evaluation with ten users already acquainted with SAT. We conducted informal interviews to find potential issues with the chatbot, with particular attention to its informativeness and user-friendliness. In light of this feedback, we adjusted our chatbot to conform to the needs of individuals unfamiliar with the SAT protocol.

\subsection{User Interface for the Study}
Our chatbot was made accessible throughout the evaluation study via a web application optimized for desktop and mobile devices running various operating systems. To further maximize accessibility, the chatbot allowed for both voice and text input and was able to give spoken and written responses. We aimed to make the application suitable for a wide variety of users and ages by allowing the choice of either `formal' or `informal' conversation, leveraging the different pools of utterances that we created for this purpose. In addition, to facilitate usability, we included in the web application the following:
(1) information about the study and introductory information about SAT, visible at login; 
(2) a page for frequently asked questions (FAQ) with in-depth explanations of the core concepts of SAT;
(3) the SAT Teacher extension, described in Section~\ref{sat-teacher-sec}, for answering any further questions about SAT and the chatbot itself on demand. Our analysis shows that over 90\% of participants used the SAT Teacher. The platform's interface is shown in Figure \ref{fig:website}.

\subsection{Experiment Setup}
We recruited N=51 volunteers among Farsi speakers from Iran and Afghanistan, the two countries with the largest Farsi-speaking populations globally, as well as from Europe, the US, and Canada. The volunteers were aged between 19 and 65 (mean = 29.46, STD = 10.97). Familiarity with the Self-Attachment protocol was not a requirement, and thus, only a fraction of our sample had engaged with SAT prior to the study. Study participants were instructed to interact with the chatbot consistently for ten days. At the end of this period, volunteers were asked to fill out a questionnaire and provide both quantitative and qualitative feedback on the chatbot's usability and performance.

\subsection{Results}
\begin{figure}
  \centering
  \includegraphics[width=0.94\textwidth]{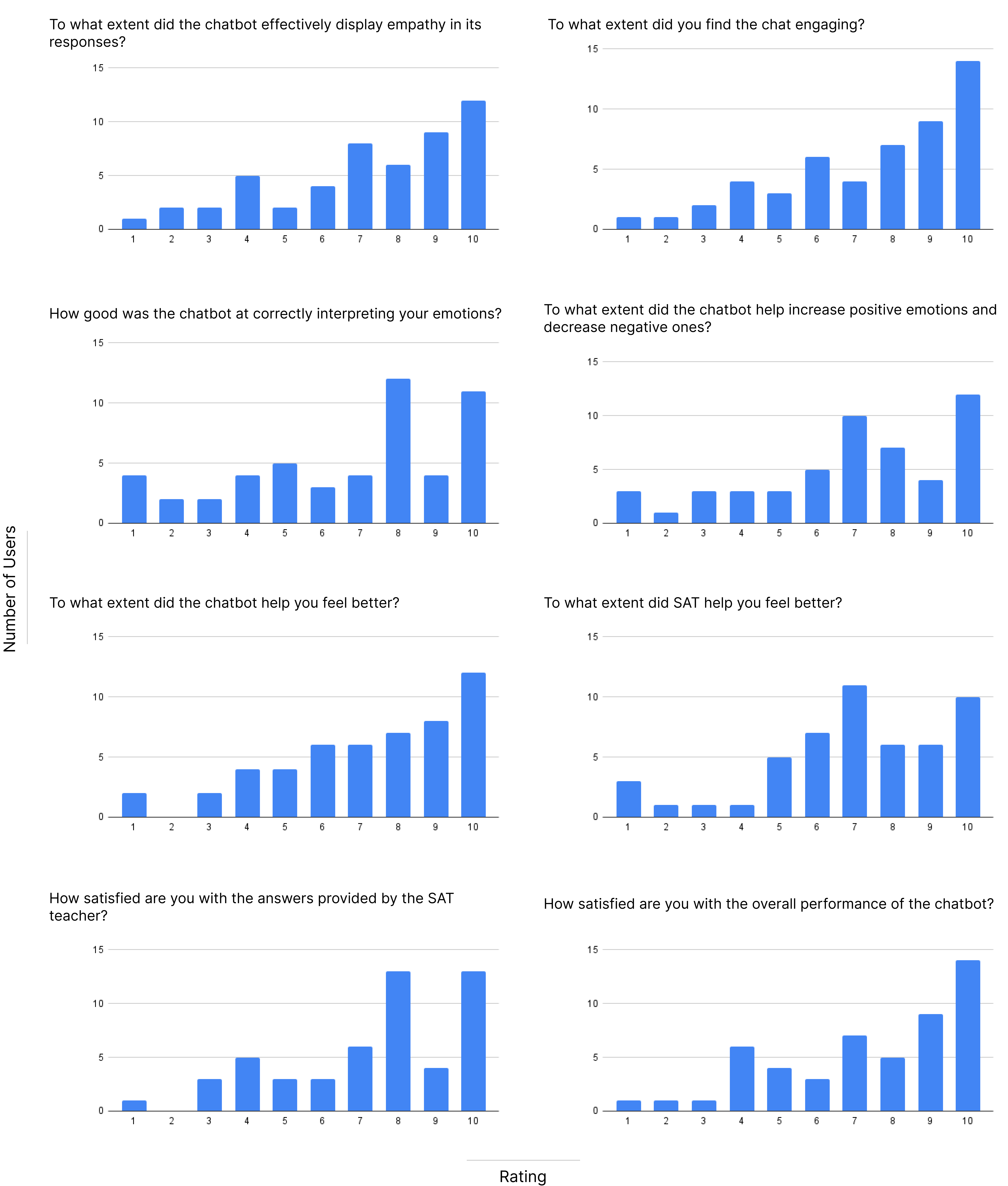}
  \caption{The chatbot's evaluation results.}
  \label{fig:results}
\end{figure}

In line with previous research \cite{9750315, 10063528}, we asked the volunteers to evaluate the chatbot's empathy, engagingness, ability to interpret human emotions correctly, and overall performance. Additionally, we asked them to score the effectiveness of the chatbot at increasing positive emotions, decreasing negative ones, and helping users feel better, as well as the effect of SAT on overall well-being. Lastly, we evaluated users' satisfaction with the answers provided by the SAT teacher. For all of the above metrics, study participants were asked to assign a discrete score between 1 and 10, where 10 indicated strongest performance. The results of this evaluation are shown in Table \ref{tab:results}, and a comprehensive breakdown of the scores assigned by the volunteers for each question is given in Figure \ref{fig:results}.

In the interest of carrying out a thorough qualitative evaluation as well as a quantitative one, we also collected open-ended user feedback on the chatbot's main benefits and limitations. We then clustered these comments in a thematic analysis into 13 overarching themes, shown in Table \ref{tab:qualitative} and accompanied by a representative feedback sample that we manually selected.

\begin{table}[!t]
\centering
\caption{Results of the quantitative evaluation.}

\renewcommand{\arraystretch}{1.1}
\begin{tabular}{lcc}
\toprule
{} & \textbf{\small{Mean score (out of 10)}}   & \textbf{\small{STD}} \\
\cmidrule{2-3}

\small{To what extent did the chatbot effectively display empathy in its responses?} & \small{7.34} &  \small{2.51} \\ 

\small{To what extent did you find the chat engaging?} & \small{7.58} &  \small{2.29} \\ 

\small{How good was the chatbot at correctly interpreting your emotions?} & \small{6.78} &  \small{2.84} \\ 

\small{To what extent did the chatbot help increase positive emotions and decrease negative ones?} & \small{7.00} &  \small{2.60} \\ 

\small{To what extent did the chatbot help you feel better?} & \small{7.34} &  \small{2.40} \\ 

\small{To what extent did SAT help you feel better?} & \small{7.06} & \small{2.50} \\ 

\small{How satisfied are you with the answers provided by the SAT teacher?} & \small{7.42} & \small{2.29} \\ 

\small{How satisfied are you with the overall performance of the chatbot?} & \small{7.48} &  \small{2.40} \\  \bottomrule

\end{tabular}
\label{tab:results}
\end{table}

\begin{table}[!t]
\centering
\caption{Results of the qualitative evaluation.}

\setlength{\tabcolsep}{0.5pt}
\renewcommand{\arraystretch}{1.4}
\begin{tabular}{lcc}
\toprule
\textbf{\small{Overall theme}} & \textbf{\small{Cluster size}} & \textbf{\small{Representative sample}} \\ \midrule
\small{Helpful} 
& \small{16} & \makecell{\small{Its empathy and willingness to help me were fascinating. The proposed} \\ \small{techniques were quite helpful.}} \\ 

\small{User-friendly} 
& \small{10} & \small{I feel the platform was more intuitive and user-friendly than the English version.} \\

\small{Beneficial exercises}
& \small{9} & \small{The suggested exercises helped me significantly.} \\

\small{Lack of open-endedness} 
& \small{8} & \small{It would be better to be able to talk to the chatbot more conveniently with open text.} \\ 

\small{Useful SAT Teacher module} 
& \small{5} & \small{I found SAT Teacher very useful for becoming familiar with SAT.} \\ 

\small{Difficult exercises} 
& \small{5} & \small{Some exercises were hard to do.} \\ 

\small{Rigid answers} 
& \small{4} & \makecell{\small{The answers were not customized. Seemingly, they were mainly}
\\ \small{following some predefined templates and patterns.}} \\ 

\small{Accessible} 
& \small{4} & \small{These bots can be really accessible for people to improve their feelings.} \\ 

\small{Insufficient guidance} 
& \small{4} & \small{Exercise descriptions were not completely informative.} \\ 

\small{Appropriate tone} 
& \small{3} & \small{I liked the feature of talking to the chatbot in both formal and friendly manners.} \\ 

\small{Delayed responsiveness} 
& \small{3}  & \small{The chatbot was a bit slow.} \\ 

\small{Lack of images} 
& \small{3} & \small{You should add some images to the suggested exercises.} \\ 

\small{Lack of empathy}
& \small{3} & \small{People usually need some empathy at the beginning of a conversation.} \\ 

\midrule
\end{tabular}
\label{tab:qualitative}
\end{table}

%%%%%%%%%%%%%%%%%%%%%%%%%   Discussion   %%%%%%%%%%%%%%%%%%%%%%%%%%%
\section{Discussion}

In this section, we discuss our contributions and findings, we examine the limitations of the study, and we make recommendations for potential future work directions.

\subsection{Theoretical and Practical Contributions}

Firstly, to the best of our knowledge, this is the first psychotherapy chatbot in Farsi and also the first in a language other than Hebrew in the Middle East, whose populations have experienced decades of mass trauma. To maximize the accessibility of our platform, we made it available online for both desktop and mobile devices and ensured that any interaction with the chatbot could support both audio (talking and listening) and visual (reading and writing) modalities. Additionally, our platform is suitable for users already familiar with the SAT protocol as well as those who have no prior experience with it, thanks to the presence of the SAT Teacher module. We assessed all aspects of our platform and investigated its strengths and limitations through a user study with N=51 participants who provided quantitative and qualitative evaluations.

Secondly, we have collected and shared three Farsi datasets with the broader research community, addressing the relatively low-resource nature of Farsi in NLP research \cite{strassel-tracey-2016-lorelei} and the previously noted difficulty of finding good-quality monolingual datasets in Farsi~\cite{shamsfard2019challenges, arshia2022peqa}. Our datasets consist of (1) A corpus of 152,000 chatbot utterances in Farsi suitable for a mental healthcare context; (2) A dataset of 6,168 Farsi utterances labeled with 12 emotions; (3) 1,100 utterances in Farsi labeled by three annotators with their perceived degree of empathy on a discrete scale from 0 to 2.

Lastly, not only does the introduction of a Farsi chatbot aid psychotherapy accessibility for people who lack access to in-person mental healthcare services in Farsi-speaking countries, but it also opens up possibilities for developing similar technologies in other languages, leveraging pre-trained models, dataset enhancement through translation, and model reproducibility with analogous architectures.

\subsection{Findings}

The quantitative feedback collected during the study indicates that the majority of volunteers scored the chatbot positively, with very similar average scores for all evaluation dimensions, as evidenced in Table \ref{tab:results}. Among the best scoring features were the chatbot's engagingness (mean = 7.58,  STD = 2.29), the overall performance (mean = 7.48, STD = 2.40), and the SAT Teacher's responses (mean = 7.42, STD = 2.29). The accuracy of the emotion recognition module, while receiving slightly lower scores than other metrics (mean = 6.78,  STD = 2.84), was still deemed adequate by most participants. While largely positive, these scores, ranging between 6.78 and 7.58 on a scale from 1 to 10, show that the platform could be further improved. Indeed, we found that while the majority of study participants rated the platform positively (i.e., with a score between 6 and 10), approximately a quarter assigned a low score (i.e., a score between 1 and 5) to each point in the evaluation questionnaire, as evidenced in Figure \ref{fig:results}.

Among the qualitative feedback (Table \ref{tab:qualitative}), the helpfulness of the chatbot was most often mentioned (16 comments), followed by its user-friendliness (10 comments) and the beneficial effects of the recommended SAT exercises (nine comments). While negative comments were overall less frequent than positive ones, several users disliked the semi-structured nature of the conversation and stated they would prefer to be able to talk to the chatbot in a completely open-ended manner (eight comments). Other, less frequent themes that may have negatively impacted the perception of our platform revolved around (1) the difficulty of some of the SAT exercises (noted by five users), (2) the rigidity of the chatbot's responses (noted by four users); (3) insufficient information and guidance to successfully complete the SAT exercises (noted by four users); (4) the perceived slowness of the platform (noted by three users); (5) lack of images that could otherwise make the material more engaging, particularly in exercise descriptions (noted by three users); (6) not enough empathy toward the user throughout the conversation (noted by three users).

%%%%%%%%%%%%%%%%%%%%%%%%%%   Limitations and Future Work   %%%%%%%%%%%%%%%%%%%%%%%%%%%
\subsection{Limitations and Future Work}
The work presented in this paper focuses on a retrieval-based chatbot that delivers digital psychotherapy in Farsi by augmenting a structured, rule-based flowchart with diverse responses fetched from a large pool. Generally, the rule-based approach is still regarded as the safest in mental health settings \cite{HARRER2023104512}, and thus it is widely used \cite{woebot-generative}. On the other hand, it is also more repetitive and potentially less engaging than fully generative methods \cite{boringbots}. In the era of large language models (LLMs), it would be possible to engineer a similar application that relies on an LLM \cite{farhat2023chatgpt}. It should be noted, however, that LLMs are known to suffer from hallucination \cite{10.1145/3571730} and can produce toxic and harmful speech \cite{zhuo2023red}. Therefore, we consider them currently unsuitable for a mental health context.

There are, however, techniques other than LLMs to achieve open-ended dialogue. For example, smaller transformer models can be finetuned on safe in-domain data \cite{rashkin-etal-2019}, and rule-based and generative techniques can be combined together to achieve greater flexibility while preventing the conversation from derailing into unwanted territory \cite{10.1145/3357384.3357881}. Future work could further explore these approaches and their effect within the context of the Farsi language.

We note that in previous 8-week-long SAT interventions, participants had access to a taught course to familiarise themselves with the exercises \cite{edalat2023VRSAT}. In contrast, in our feasibility study, only written materials and the SAT Teacher module were available to provide information about SAT. Additional teaching methods, such as informational videos about SAT and its practice, could eliminate any negative feedback derived from the exercises' difficulty or the clarity of their directions. 

Lastly, ten days may not be enough to evaluate all aspects of the chatbot, e.g., its potential repetitiveness over an extended period of time or its long-term effect on mental health. Future studies should ideally cover a longer time frame.

%%%%%%%%%%%%%%%%%%%%%%%%%%   Conclusion   %%%%%%%%%%%%%%%%%%%%%%%%%%%
\section{Conclusion}
We presented our work on a Farsi chatbot to guide users through the Self-Attachment Technique (SAT), and we collected and shared novel datasets in Farsi to aid future work in this area of research. Through human evaluation, we observed that the chatbot was able to converse in an engaging manner with the users and recommend helpful SAT exercises. Overall, our feasibility study shows promise in applying chatbot technologies to supplement the delivery of psychotherapeutic interventions in the context of low-resource languages.

\clearpage

\bibliographystyle{ACM-Reference-Format}
\bibliography{sample-manuscript}

\end{document}